\documentclass{IEEEtran}
\usepackage{cite}
\usepackage{amsmath,amssymb,amsfonts}
\usepackage{graphicx}
\usepackage{textcomp,nicefrac}

\usepackage{soul}
\usepackage[bookmarks=false]{hyperref}
\usepackage[center]{caption}
\usepackage{tikz,tikz-qtree}
\usepackage{listings}
\graphicspath{{./figures/}}
\newcommand{\osquare}{$O^2$ }

\def\BibTeX{{\rm B\kern-.05em{\sc i\kern-.025em b}\kern-.08em
T\kern-.1667em\lower.7ex\hbox{E}\kern-.125emX}}
\begin{document}
\title{The ReadoutCard userspace driver for the new Alice $O^2$ computing system}
\author{Konstantinos Alexopoulos and Filippo Costa on behalf of the ALICE
  collaboration
\thanks{Manuscript submitted October 30, 2020.}
\thanks{K. Alexopoulos is with the European Organization for Nuclear Research,
CERN, CH-1221 Geneva, Switzerland (email: kostas.alexopoulos@cern.ch)}
\thanks{F. Costa is with the European Organization for Nuclear Research, CERN,
  CH-1221 Geneva, Switzerland (email: filippo.costa@cern.ch)}}

\maketitle

\begin{abstract}
The ALICE (A Large Ion Collider Experiment) experiment focuses on the study of
the quark-gluon plasma as a product of heavy-ion collisions at the CERN LHC
(Large Hadron Collider). During the Long Shutdown 2 of the LHC in 2019-2020, a
major upgrade is underway in order to cope with a hundredfold input data rate
increase with peaks of up to 3.5 TB/s. This upgrade includes the new
Online-Offline computing system called O$^2$. 

The O$^2$ readout chain runs on commodity Linux servers equipped with custom
PCIe FPGA-based readout cards; the PCIe v3 x16, Intel Arria 10-based CRU (Common
Readout Unit) and the PCIe v2 x8, Xilinx Vertex 6-based CRORC (Common ReadOut
Receiver Card). Access to the cards is provided through the O$^2$ ReadoutCard
userspace driver which handles synchronization and communication for DMA
transfers, provides BAR access, and facilitates card configuration and
monitoring. The ReadoutCard driver is the lowest-level interface to the readout
cards within O$^2$ and is in use by all central systems and detector teams of
the ALICE experiment.

This communication presents the architecture of the driver, and the suite of
tools used for card configuration and monitoring. It also discusses its
interaction with the tangent subsystems within the O$^2$ framework.
\end{abstract}


\section{Introduction}
\label{sec:introduction}
\IEEEPARstart{A}{LICE} is an experiment at the CERN LHC (Large Hadron Collider)
focusing on the study of the quark-gluon plasma - a state of matter which
existed shortly after the Big Bang - as a product of heavy-ion collisions.
Currently, the second long shutdown (LS2) of the LHC is underway, allowing for
preparations for Run3, which will run at significantly higher luminosity. The
main physics topics addressed by the upgrade require measurements characterized
by a very low signal-over-background ratio, making traditional triggering
strategies inefficient. Hence, the Time Projection Chamber (TPC) necessitates
the implementation of continuous readout, capable to keep up with an interaction
rate of 50kHz \cite{b_loi}. During LS2, ALICE is upgrading its detector and
software systems to fulfill the above requirements and achieve a higher
resolution.

\subsection{The O$^2$ Computing System}
The upgrade includes the new Online-Offline computing system which is called
\osquare \cite{b_o2}, extending across two major computing slices. The
$O^2$/FLP subsystem is comprised of ~200 First Level Processors (FLPs)
responsible for detector readout. They are equipped with specialized data
acquisition cards that interface with the Front-End Electronics (FEE) of the
detectors. FLPs are designed to handle triggered and continuous readout
operation, without discarding any events. A first-level grouping of the read-out
events takes place within the FLPs before the data flow to the next subsystem.
$O^2$/FLP also includes the quality control system and services for control,
configuration, monitoring, logging and bookkeeping. The $O^2$/EPN subsystem
is comprised of ~250 Event Processing Nodes (EPNs) which are responsible for
synchronous calibration and reconstruction before the data reaches storage.

For what concerns \osquare the data originate from the Readout Cards of the
experiment, namely the Common ReadOut Receiver Card (CRORC) and the Common
Readout Unit (CRU), which are connected to the Front-End Cards of the detector
electronics. In this document we focus on the ReadoutCard userspace driver,
which controls and provides a communication interface to the cards of the
experiment, serving as the lowest-level interface to them within the \osquare
framework.

The ReadoutCard package was initially published in \cite{b_roc_og}. Since then
the userspace driver has been heavily developed to improve implementation
details and extend core functionality. The existing library interfaces have been
improved and fragmented to limit their functionality to their expected uses,
reducing unnecessary complexity. Moreover, several interfaces have been
introduced to facilitate access to new software components. The available tools
have been extended to provide a complete suite to configure and monitor the
status of the cards. Finally, ReadoutCard has been integrated with the \osquare
Monitoring \cite{b_mon} and \osquare Configuration \cite{b_conf} components.
This paper presents the state of ReadoutCard following these developments.

\section{Hardware}

The readout chain is designed around the two readout cards of the experiments.

\subsection{CRORC} 
\begin{figure}[ht]
  \begin{minipage}[t]{0.49\linewidth}
    \centering
    \includegraphics[width=1\textwidth]{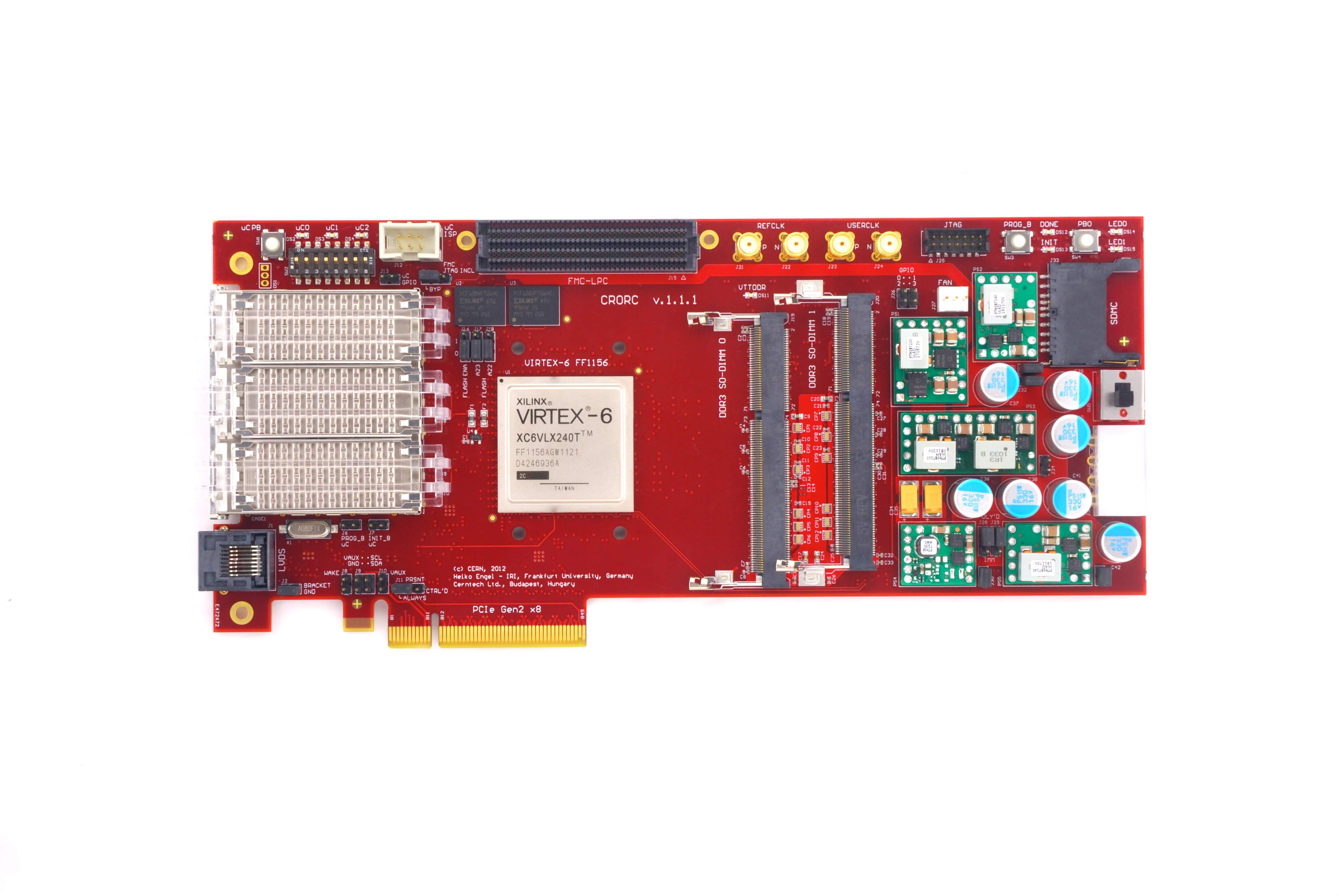}
    \caption{CRORC}
    \label{crorc}
  \end{minipage}
  \begin{minipage}[t]{0.40\linewidth}
    \centering
    \includegraphics[width=1\textwidth]{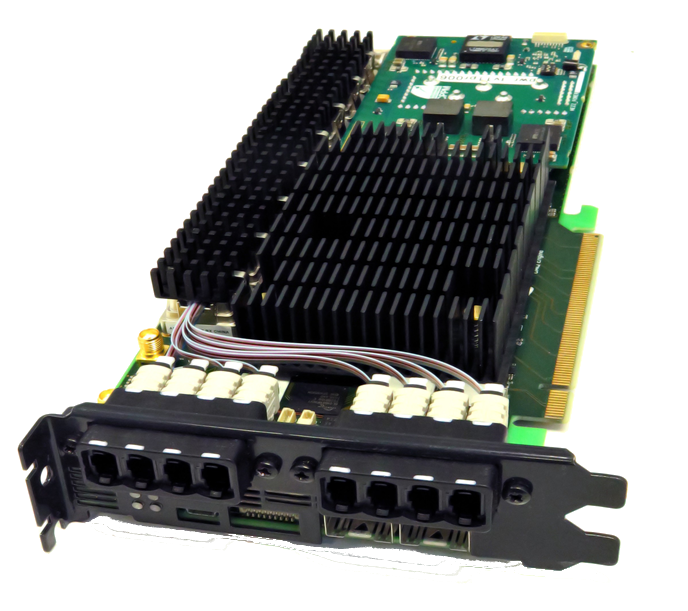}
    \caption{CRU}
    \label{cru}
  \end{minipage}
\end{figure}

The Common ReadOut Receiver Card \ref{crorc} was used to
collect data from the majority of detectors of the ALICE experiment during Run2
\cite{b_crorc}. It is a PCIe v2 x8 card based on the Xilinx Vertex 6 FPGA,
equipped with 12 optical links able to run up to 6 Gb/s each.

The CRORC uses the Detector Data Link (DDL) protocol \cite{b_ddl}. A big part of
its old firmware has been reused and extended so that it can address the new
needs of the detectors and the \osquare facility during Run3. This enables the
reuse of DDL-enabled hardware from Run2 for a few detectors that would not
benefit from higher throughput capabilities.

\subsection{CRU}

The Common Readout Unit \ref{cru} is the main readout card for the ALICE experiment during
Run3. It is based on the PCIe40 \cite{b_pcie40} hardware designed for LHCb, a PCIe v3 x16 card
equipped with the Intel Arria 10 FPGA. Connection to the FEE of the detectors
happens via up to 24 optical fibers, which can be used for readout, trigger,
timing and/or slow control, depending on individual needs. 

The CRU uses the GBT protocol \cite{b_gbt} for its readout links and its firmware is under
active development \cite{b_cru}.

\subsection{FLPs}

Following an extensive software and hardware assessment process
\cite{b_flp_ass}, and a competitive tender, the DELL PowerEdge R740 servers were
selected to run the FLP portion of the \osquare computing facility. The FLPs
come in two flavors, silver and gold, depending on computing needs for data
processing. Both flavors run CERN CentOS 7 and are equipped with 96GB of DDR4
2666MT/s RAM and a 480GB SSD @ 6 Gbps. The Silver version uses 2 Intel Xeon
Silver Cascade-Lake 4210s, each with 10 cores @ 2.2 GHz, whereas the Gold
version uses 2 Intel Xeon Gold Cascade-Lake 6230s, each with 20 cores @ 2.1 GHz.
A single server may be a equipped with up to four CRORCs, or up to three CRUs,
depending on detector and FEE topology and configuration.

\section{Kernel Driver}
The first layer over the PCIe interface to the cards is the PDA (Portable Driver
Architecture) UIO (Userspace IO) kernel module \cite{b_pda_thesis}, developed by the Frankfurt Institute for
Advanced Studies (FIAS). PDA also provides a userspace library in C \cite{b_pda}
which supports device enumeration and provides a handle to PCIe devices.

Through the PDA handle the following functionality may be exploited:
\begin{itemize}
  \item Registering DMA memory targets with the IOMMU (Input Output Memory
    Mapping Unit), which maps the PCIe physical address space to the virtual
    address space of the CPU. This allows the mapping of separate memory regions
    to a contiguous memory space, while also preventing invalid memory accesses.
  \item If the IOMMU component is not physically present in the system or is
    inactivated, PDA generates DMA scatter-gather lists in order to ensure
    access to the non-consecutive physical memory spaces of the DMA engine of
    the device.
  \item Memory mapping the BAR (Base Address Registers) of the device to a
    virtual address space. This allows the execution of BAR operations on the
    process level to be carried out by means of simple memory reads and writes.
\end{itemize}

\section{Userspace Driver}

ReadoutCard \cite{b_roc_gh} is a C++ userspace driver and library, which wraps around PDA
functionality to access and control the cards. Both cards of the experiment, the
CRORC and the CRU, are fully supported by the ReadoutCard package, and access is
published through a BAR interface as well as a high-level DMA channel interface.

\begin{table}[h]
  \begin{center}
\caption{CRU/CRORC Access Interfaces}
\label{tab:cru_crorc_acc}
\setlength{\tabcolsep}{3pt}
    \scalebox{1.4}{
\begin{tabular}{|c||c|c||c|c|c|c|c|c|}
\hline
Card&
\multicolumn{2}{c||}{CRU}&
\multicolumn{6}{c|}{CRORC}\\
\hline
Endpoint&
0& 1&
\multicolumn{6}{c|}{0}\\
\hline
Link \#&
0-11& 12-23&
0& 1& 2& 3& 4& 5\\
\hline
BAR \#&
0/2& 0/2&
0& 1& 2& 3& 4& 5\\
\hline
DMA Channel&
0& 1&
0& 1& 2& 3& 4& 5\\
\hline
\end{tabular}}
\label{tab:cru_crorc_acc}
\end{center}
\end{table}

\subsection{Addressing}

ReadoutCard accesses the cards on the level of an \textit{endpoint}, which
coincides with a PCIe endpoint.

The CRORC has a total of 6 optical connections, all within a single endpoint,
and for each of them a BAR handle can be acquired and a DMA channel opened. In
other words, the access granularity level of the CRORC is a link, for all
interfaces.

The CRU boasts 24 optical connections, with each physical card separated to two
logical \textit{endpoints}. Each endpoint owns 12 of the 24 links.  Contrary to
the CRORC, the CRU follows a different scheme to access its links.  It publishes
2 BARs per endpoint: BAR 0 publishes DMA related registers and facilitates DMA
orchestration on the endpoint level, whereas BAR 2 publishes the rest of the
registers, i.e.  configuration, monitoring, etc. For what concerns the DMA
functionality of the CRUs links, each endpoint orchestrates the DMA transfer
through a single DMA channel.

Table \ref{tab:cru_crorc_acc} summarizes the addressing details that are
specific to each card.

ReadoutCard offers multiple ways of addressing an endpoint: 
\begin{itemize}
  \item Using the \textbf{PCI address} comprised of bus, device, and function
    number assigned to the PCI device (e.g. \lstinline{3b:00.0}).
  \item Using the \textbf{Serial ID} of the device, a pair of the serial number
    of the card, which is unique to every physical card, and its endpoint (0 or
    1) (e.g. \lstinline{0233:1}).
  \item Using the \textbf{Sequence ID} of the endpoint, a sequential number
    assigned by ReadoutCard aiming to simplify development and testing (e.g.
    \#\lstinline{2}).
\end{itemize}

\subsection{Enumeration}
For what concerns the PDA driver, PCIe devices are enumerated on kernel module
insert. When ReadoutCard requests a handle it receives a handle to a PCI Device,
which can, by means of the Vendor and Device ID, be classified as a CRORC or a
CRU. Upon receiving the device handle by PDA, ReadoutCard builds upon it to
create a \lstinline{RocPciDevice} object. This object holds a reference to the
PDA device handle, initialized handles to the BAR interface(s) of the device, as
well as a \lstinline{CardDescriptor} struct. The latter serves as a standalone
identifier for a given endpoint, holding addressing information.

The handle acquisition process takes place every time a process requests a device
scan, or an interface to a specified device through the ReadoutCard library.
Information gathered during the initial stages remains available throughout the
lifetime of the process, allowing for an efficient transition between devices
and interfaces.

\subsection{Compatibility}

The driver is accessing the card at the lowest level through its published
registers in the BAR. Between firmware versions the functionality of a given set
of registers may change, ranging from a simple shift of control bits to a
completely different sub-module. As a result, a firmware incompatibility between
what is expected by the driver, and what is used by the device, can lead to an
unexpected state within the device or simply faux operations.

To address this, ReadoutCard checks the firmware of the device in question
against a list of supported versions. This check is implemented on the level of
the DMA interface as well as for the various command line tools which need to be
protected. In all cases an option to bypass the check is available, which may be
necessary during development and testing in between releases.

\subsection{Synchronization}

For its synchronization needs, ReadoutCard employs an internal mutex,
which uses an abstract UNIX socket as its underlying locking mechanism, called a
\lstinline{SocketLock}. The \lstinline{SocketLock} attempts to bind to the UNIX
socket with the specified name, and if it is successful it keeps the connection
open, thus guaranteeing atomic access to the underlying resource, in this case
PDA access.

As an example, the PDA library is not thread-safe. It is necessary to ensure
that exclusive access is granted whenever a PDA operation is taking place,
otherwise the kernel module or the device might end up in an unsafe state, with
undefined side-effects. 

Even though the \lstinline{SocketLock} is not the most performant among modern
interprocess synchronization mechanisms (see boost::interprocess \cite{b_bip}),
securing access for the needs of ReadoutCard takes a small portion of the
execution time which leaves throughput-dependent sections unaffected. Moreover,
robustness is much more important, as it is dictated by the large number of
users and processes that are active on an FLP at any time. The
\lstinline{SocketLock} guarantees that in case of a pathological scenario (e.g.
a process crash), resources will be automatically released, eliminating the need
for any manual intervention.

\subsection{BAR Interface}
ReadoutCard utilizes PDA to request a map of the BAR of the device in its
virtual memory space. Consequently ReadoutCard implements operations for BAR
reads and writes by indexing its BAR-mapped virtual address space. By performing
basic BAR limits checks it ensures that only legal BAR accesses are carried out.

BAR accesses are used for all communication with a device, be that data-taking
orchestration, configuration, or monitoring. BAR communication needs to be
exclusive during data-taking and, depending on the sub-component, during
configuration. Atomicity during data-taking is anyway guaranteed as will be
discussed in the next section, whereas atomicity during configuration is assured
through tangent packages on a case-by-case basis.

As a result the BAR interface does not itself enforce mutual exclusion. This is
vital to ensure low-latency and service continuity, as multiple tools and users
need to constantly communicate with the cards in a responsive way. The number of
concurrent processes accessing the BAR interface of a card is only limited by
the resources of the system.

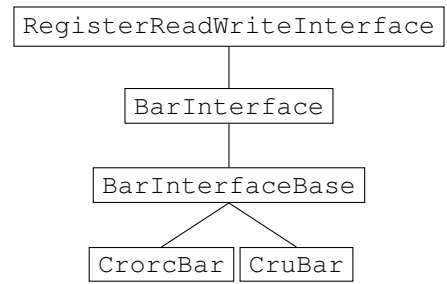
\begin{figure}[h]
\centering
\begin{tikzpicture}[every tree node/.style=draw]
\tikzset{level distance=30pt}
\Tree [.{\lstinline{RegisterReadWriteInterface}}
  [.{\lstinline{BarInterface}}
  [.{\lstinline{BarInterfaceBase}}
  [.{\lstinline{CrorcBar}} ]
  [.{\lstinline{CruBar}} ]
  ] ] ]
\end{tikzpicture}
\caption{BAR class hierarchy}
\label{fig_bar_imp}
\end{figure}

\subsubsection*{Implementation}

BAR accesses rely on two interfaces. The \lstinline{RegisterReadWriteInterface}
defines functions for reading, writing, and modifying a register. The
\lstinline{BarInterface} defines functions that utilize the BAR and are common
between the CRU and the CRORC. The \lstinline{BarInterfaceBase} initializes the
device if needed, maps the BAR address space to virtual memory, and implements
the \lstinline{RegisterReadWriteInterface}. The \lstinline{BarInterfaceBase} is
extended for every card, so as to implement all card-specific operations,
resulting in the \lstinline{CrorcBar} and \lstinline{CruBar} classes, as seen in
figure \ref{fig_bar_imp}.

\subsection{DMA}
DMA is used for data acquisition, which presents the heaviest requirements in
terms of throughput. A DMA transfer is facilitated through the use of a DMA
channel, which is uniquely opened on the level of an endpoint.

Initializing a DMA channel involves a handshake procedure between the card and
the driver, which makes sure that the card is properly reset and relevant
structures on both sides like buffers, FIFOs, and counters are initiated in order
to support data-taking. Moreover, a userspace shared memory buffer is registered
with PDA as the DMA buffer of the transfer, which is in turn validated by
ReadoutCard.

Starting with this process, mutual exclusion needs to be guaranteed, as
concurrent actions on the same channel will certainly have unintended
consequences, invalidating the conditions of the ongoing transfer and
overwriting contended buffer memory, leading to a data-taking halt or corrupted
data.

To mitigate this, ReadoutCard once again exploits \lstinline{SocketLock}
functionality, acquiring the mutex before any hand-shaking takes place, and
holding it until the data-taking is finished and relevant buffers have been
purged. As a consequence, concurrent attempts to open a DMA channel fail
gracefully.

\subsubsection{Memory Layout}

\begin{figure}[h]
\centering
\includegraphics[width=0.8\columnwidth]{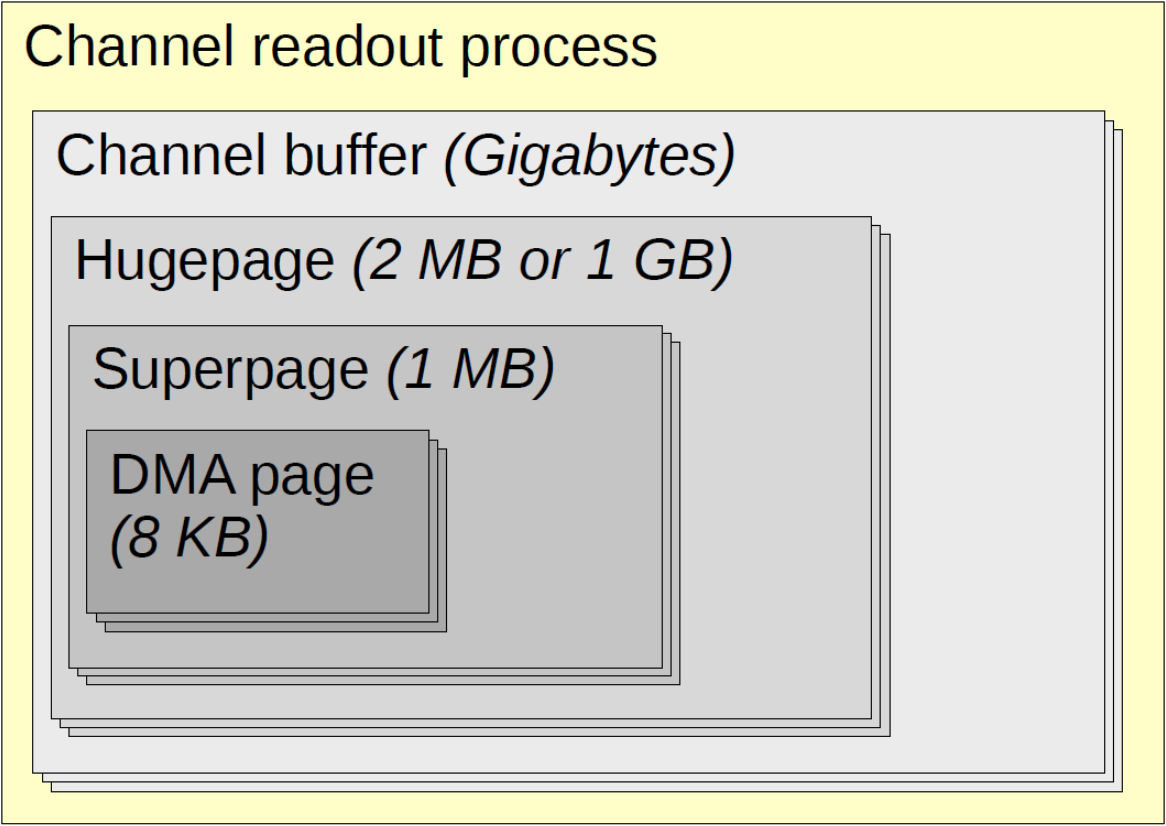}
\caption{Memory Layout (typical sizes)}
\label{dataformat}
\end{figure}

DMA transfers are performed using DMA channel buffers, made up of shared memory,
and are comprised of Superpages. A Superpage (usually 1 MiB) is the granularity
level on which the driver and the cards communicate. Each Superpage contains DMA
pages, which have a variable size of up to 8KiB, even within the same Superpage
\ref{dataformat}. Every DMA page includes a header followed by the payload.
With regard to its DMA functionality, the driver is agnostic to the content and
structure of a Superpage.

\subsubsection{Communication}

As discussed before, DMA channel functionality is always wrapped in a
\lstinline{SocketLock} acquisition associated with the endpoint in question,
which guarantees the mutual exclusion property. It is also necessary to clean up
potentially leftover PDA buffers as a result of past process crashes. This
action also happens under the protection of a \lstinline{SocketLock}, albeit a
different one, the one that is tied to PDA communication.

In order to orchestrate the DMA transfer, the DMA channel holds open interfaces
to the BAR, or BARs, of the card, and also publishes convenience functions to
monitor health metrics of the transaction, such as the number of dropped packets
or the FPGA temperature of the card.

On the successful outcome of the above, DMA buffer registration needs to take
place. ReadoutCard provides the PDA driver with a pointer to a userspace buffer
to be registered for DMA. For performance reasons, this buffer should be as
physically contiguous as possible.  To address this, ReadoutCard supports
hugepage-backed buffers. HugePage support is a Linux kernel feature that enables
OS support of memory pages larger than the default (usually 4K). ReadoutCard
uses 1GiB hugepages, greatly limiting the number of DMA-related memory pages. As
a consequence, memory page swaps to disk and page table entry lookup times are
greatly reduced. HugePages have been an asset that was heavily used during
development and testing. However, ReadoutCard will accept any shared memory
buffer as long as it is contiguous. In any other case the IOMMU needs to be
enabled, otherwise an exception will be thrown.

After PDA initializations, DMA buffer registration, and some exchange of
information between ReadoutCard and the underlying card have concluded,
data-taking may commence. For both CRORC and CRU the underlying mechanism
enabling DMA transfers is largely the same and both cards are ``superpage
aware''.  The ReadoutCard functionality with regard to DMA transfers boils down
to Superpage address and size information exchange, which is facilitated through
the use of two Superpage queues. 

\begin{figure}[h!]
\centering
\includegraphics[width=1\columnwidth]{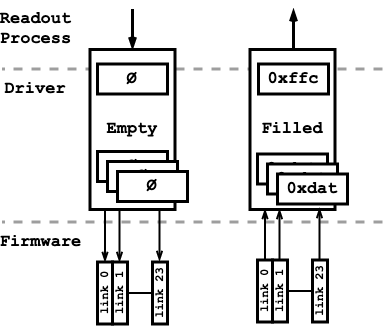}
\caption{Superpage Data Flow}
\label{fig_dataflow}
\end{figure}

The first one is called \lstinline{TransferQueue} and manages the ``free''
Superpages, i.e the Superpages that are empty and which may be populated by the
card. When such a Superpage becomes available in the queue, ReadoutCard checks
the FIFOs of the enabled links of the card for free slots, and proceeds to push
the Superpage, or Superpages, in a round-robin fashion between the links. A
Superpage push consists of translating the address of the free Superpage, from
the virtual address space of the process to the bus address space, before
placing it and its size in the relevant FIFOs and notifying the firmware through
the BAR.

When a Superpage is filled, or ``ready'', the firmware notifies ReadoutCard by
means of an incrementing per-link counter, which will in turn transfer this
Superpage to the second queue, the \lstinline{ReadyQueue}. Superpages in this
queue are already filled with data and are ready to be read out by the relevant
process. \ref{fig_dataflow}.

\subsubsection{Implementation}

The functionality of a DMA channel is dictated by the
\lstinline{DmaChannelInterface}, which includes interfaces for controlling the
DMA state, and interacting with the two Superpage queues and the device.

The \lstinline{DmaChannelBase} implements methods that are common between the
card-specific classes, takes care of DMA channel locking, and includes the
logging facilities.

DMA buffer-related functions are implemented in \lstinline{DmaChannelPdaBase},
which also incorporates the high-level DMA start/stop/reset functionality by
means of a simple state machine.

Finally, the card-specific DMA Channel classes, i.e.
\lstinline{CrorcDmaChannel} and \lstinline{CruDmaChannel}, inherit the above and
implement device specific communication, namely the interface to exchange
Superpages with the cards.

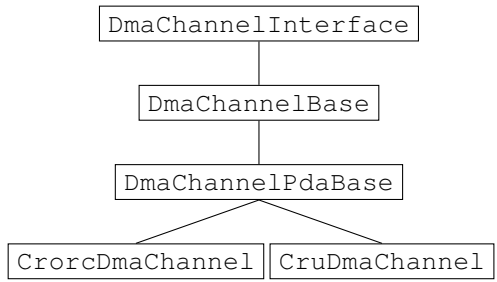
\begin{figure}[h!]
\centering
\begin{tikzpicture}[every tree node/.style=draw]
\tikzset{level distance=30pt}
\Tree [.{\lstinline{DmaChannelInterface}}
  [.{\lstinline{DmaChannelBase}}
  [.{\lstinline{DmaChannelPdaBase}}
  [.{\lstinline{CrorcDmaChannel}} ]
  [.{\lstinline{CruDmaChannel}} ]
  ] ] ]
\end{tikzpicture}
\caption{DMA class hierarchy}
\label{fig_bar_imp}
\end{figure}

\section{Software}

Apart from its userspace driver functionality, the ReadoutCard package also
publishes a library which may be primarily used for data-taking through its DMA
interface, and control, configuration, and/or monitoring through its BAR
interface.  Building upon these interfaces, ReadoutCard also provides a set of
utilities, the RoC tools, which provide complete solutions for operations that
need to be performed on the cards on both development and production contexts.

\subsection{Library}

In order to provide adequate support for the needs of the experiment,
ReadoutCard provides a number of interfaces with different scopes.

\subsubsection{DMA} A high-level C++ interface to the DMA channel of the cards,
used by processes to perform readout. This is supplied as a well defined API
that provides functions to control the DMA state by performing start and stop
operations. It also provides functions interfacing the two superpage queues of
the driver facilitating the transfer. These functions include operations to
push, pop, fill superpages and check the current status of the queues.  Moreover
the DMA channel offers some convenience functions to get information regarding
the underlying card, like its type, PCI address, firmware info, and
corresponding NUMA node, as well as to monitor health metrics of the transaction
such as the number of dropped packets and the FPGA temperature of the card.

The DMA interface is exploited by \osquare Readout \cite{b_readout}, the
high-level readout process of the experiment, to access the card, initialize,
and perform data-taking.

\subsubsection{BAR} A high-level C++ interface to the BARs of the cards, which
is heavily used for the internal needs of the driver and by external tools.
External tools normally use an interface that supports only simple BAR
operations: read, write, and modify, the \lstinline{RegisterReadWriteInterface}.
Internally however, ReadoutCard uses a derived interface, the
\lstinline{BarInterface}, which also defines functions reporting on the status
of the cards and facilitating configuration. This offers a complete view of the
underlying card for more complex use cases.

Apart from the custom detector solutions utilizing the BAR interface, it also
serves as the card communication channel for \osquare ALF (Alice Low-level
Frontend) \cite{b_alf_gh}, the process that publishes interfaces to the cards to
DCS (Detector Control System) \cite{b_dcs}, so that the latter can trigger Slow
Control operations through the former.

\subsubsection{Python BAR} A python wrapper to the BAR interface, used in
situations where rapid development cycles are paramount; mainly firmware
development and detector teams.

\subsubsection{CardConfigurator} A C++ interface to a class orchestrating
card configuration. The \lstinline{CardConfigurator} class is initialized
for a specific card with configuration parameters that may be passed
programmaticaly, or parsed from a configuration uri, as facilitated by the
\osquare Configuration library. It consequently configures all the
sub-components of the card in a modular way.

\subsubsection{Miscellaneous} ReadoutCard also offers a number of interfaces
with a more limited scope to publish software component functionality that may
be needed by tangent packages. Examples include interfaces to use the Firmware
Checker, as described earlier, control the Pattern Player, which controls the
pattern player module in the CRU, or header files containing e.g. register
addresses.

\subsection{Tools}
The CLI tools provide configuration, monitoring, and testing functionality for
the readout cards. The ones providing status output have been integrated with
the \osquare Monitoring facility which takes care of propagating the state to
the centrally managed sub-systems. Indicatively:

\begin{itemize}

  \item \textit{\lstinline{roc-list-cards}}:
    Lists the available readout cards in the system.  Provides addressing
    options and device information (e.g. firmware version, NUMA node).
  \item \textit{\lstinline{roc-config}}:
    Configures the components and links of the cards in a modular fashion. It
    takes care not to unnecessarily reconfigure modules unless explicitly forced
    to.
  \item \textit{\lstinline{roc-status}}:
    Reports card configuration status, i.e. reports the resulting state of a
    \lstinline{roc-config} execution. It also reports the status of the links,
    namely if they are UP or DOWN coupled with information regarding the optical
    connections.
  \item \textit{\lstinline{roc-metrics, roc-pkt-monitor}}:
    Report card metrics and packet statistics. These tools differ from
    \lstinline{roc-status} because they report transient information, specific to
    the current run state, like the current packet rate.
  \item \textit{\lstinline{roc-bench-dma}}:
    Performs readout and extensive error checking. Heavily used to facilitate
    development and testing cycles, especially in conjunction with the firmware
    teams.
  \item \textit{\lstinline{roc-reg-read, roc-reg-write, roc-reg-modify}}
    Perform the most basic, lowest level, register read, write, and modify operations.

\end{itemize}

\section{Performance}

Benchmarks were executed on an FLP, as described in the ``Hardware'' section,
equipped with the Silver version of the CPU and 2 CRUs.

The FLPs are equipped with dual-socket CPUs, forming two areas of locality,
where CPUs are directly attached to their own local RAM. In the NUMA
(Non-Uniform Memory Access) architecture these are called NUMA nodes. For
communication between these two areas, Intel uses the Intel Ultra Path
Interconnect (UPI) \cite{b_upi}, a low-latency coherent interconnect for
scalable multiple-processor systems in a single shared address space.

\begin{figure}[h]
\centering
\includegraphics[width=1.04\columnwidth]{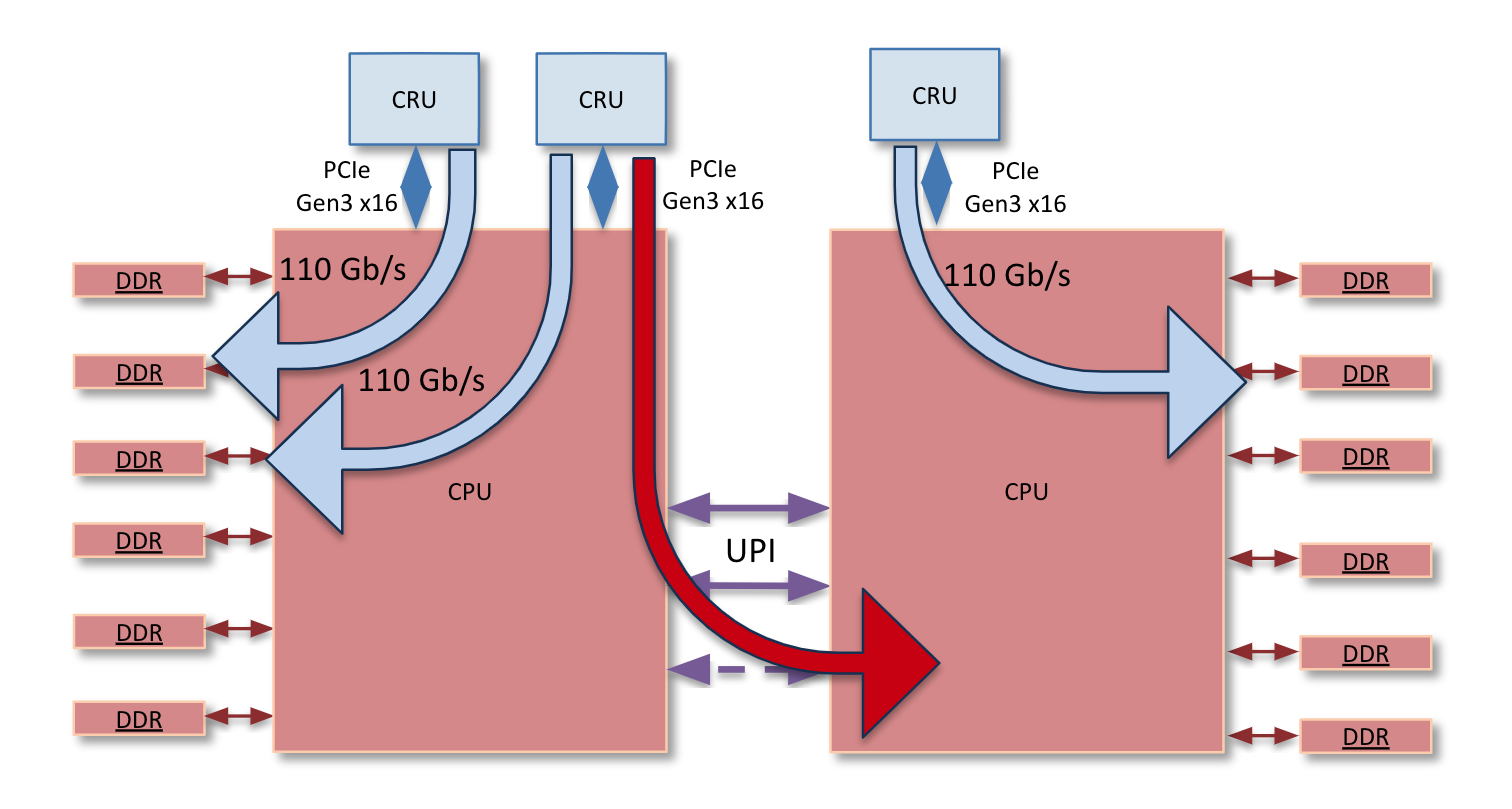}
\caption{NUMA node locality}
\label{f_upi}
\end{figure}

The readout cards installed on the FLP are connected to one of the two CPU
sockets, and thus are part of a single node. In case the process accessing the
PCIe endpoint is accessing a NUMA node that is not local to the specific readout
card, data will have to flow over the UPI link \ref{f_upi}, which for
high-throughput DMA transfers, will be a performance bottleneck. It is thus
imperative that processes are pinned to the correct NUMA node when running DMA,
a requirement which has been addressed by utilizing the \lstinline{numactl}
utility.

\subsection{DMA Performance}

\begin{figure}[h]
\centering
\includegraphics[width=1\columnwidth]{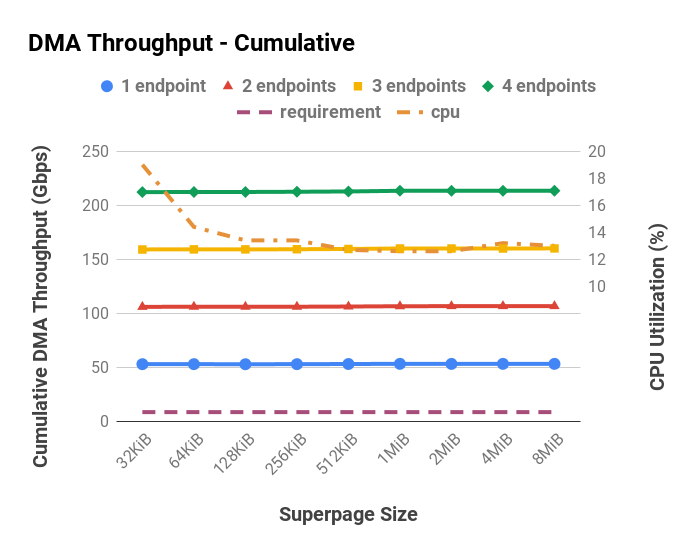}
\caption{DMA Throughput - Superpage Size - CPU Utilization}
\label{dma_bench}
\end{figure}

To measure DMA performance, each CRU was configured to use the DDG (DMA Data
Generator), an internal module generating packets. In order to maximize
throughput 8 optical links were enabled on the endpoint level. The DMA
throughput measurements are presented cumulatively with regard to endpoints used
and as a function of the superpage size. Moreover, CPU utilization is also
measured for the different scenarios. These tests were facilitated by the
\lstinline{roc-bench-dma} benchmarking tool provided by ReadoutCard. The results
can be seen in figure \ref{dma_bench}.

For a single endpoint, the DMA throughput reaches a stable 53 Gbps, which
greatly surpasses the requirement of 8.75 Gbps. In addition, the superpage size
does not in any way affect the DMA throughput, which remains stable across the
23KiB to 8 MiB range.

For all 4 endpoints, i.e. the 2 CRUs, concurrently running DMA at full speed,
the throughput scales well without any bandwidth losses to give a cumulative 212
Gbps.

Throughput remains unaffected in relation to the superpage size changes for all
endpoint layers. The same is not true however for the CPU utilization, for which
we see a sharp drop as the superpage size increases. A minimum of 13\% of
utilization is reached for sizes of 512 KiB to 2 MiB, dictating the optimum size
range to be used. Additionally, the low utilization allows for other processes
within the FLP to exploit adequate CPU resources for their needs.

\subsection{BAR Performance}

\begin{figure}[h]
\centering
\includegraphics[width=1\columnwidth]{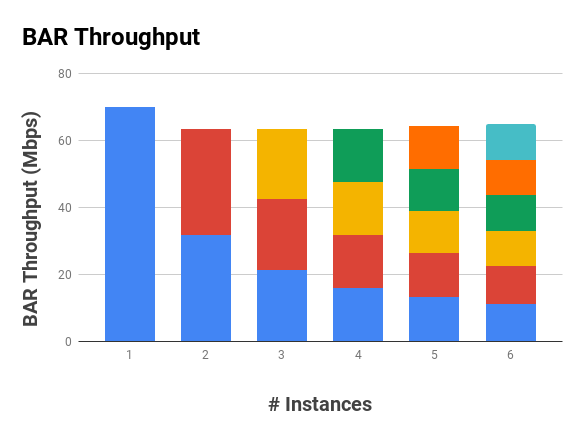}
\caption{BAR Throughput - \# Instances}
\label{bar_bench}
\end{figure}

BAR performance was measured with the \lstinline{roc-bar-stress} tool, which
runs a process accessing a single BAR interface and stressing it for a number of
operation cycles comprising of sequential read and write operations, at full
speed. For the purposes of these benchmarks a process was run for 10 million
cycles. The output of this process is equivalent to the total usable bandwidth
of a single interface. For BAR 0 the maximum throughput measured reaches 80
Mbps, or around 2.5 million 32-bit operations per seconds. For BAR 2 the maximum
throughput is slightly lower at 70Mbps, which roughly translates to around
2 million 32-bit operations per second. This inconsistency between the two BAR
interfaces is due to the fact that they are connected to firmware components
which operate on different clocks and thus operate at different speeds. BAR
performance is consistently exceeding the operational requirements of the
system.

BAR 0 is used for DMA, and under normal circumstances should only be accessed by
a single process. For BAR 2 on the other hand multiple concurrent processes are
expected to run at the FLP at any time, including monitoring daemons as well as
user-initiated operations. Consequently, it is important that the bandwidth of
the BAR is distributed fairly among all processes and no starvation is observed.
This is investigated in \ref{bar_bench}, where increasing the number of
instances accessing BAR 2 does not result in a drop of the cumulative throughput
of the interface. Moreover, available bandwidth is demonstrably distributed
fairly among all actors.

\section{Conclusion}

ReadoutCard is a userspace driver controlling the two cards of the ALICE
\osquare computing system. It provides a library with abstract interfaces to
access the cards as well as a suite of command-line tools which publish
functionality and facilitate development. The ReadoutCard package is the very
first layer to access the readout cards within the \osquare facility and has
been in heavy use by all sub-detectors and sub-systems of the experiment for the
last 3 years.

Moving towards operations the ReadoutCard package will continue to adapt to the
evolving requirements. Additionally, the suite of tools will be extended to
provide more options for debugging both during the detector commissioning phase
and production. On the software side, improvements are due with regard to
integration with other \osquare software components, specifically the ones
facilitating Logging \cite{b_il} and Control \cite{b_control}. Lastly, an effort
to improve the Continuous Integration for the package is foreseen, which will
also cover scenarios for the various detector use cases.

\end{document}